\documentclass[preprint2]{aastex}

\begin{document}
\title{The Cepheid Impostor HD 18391 and its Anonymous Parent Cluster}

\author{David G. Turner\altaffilmark{1,3,4}, V.~V. Kovtyukh\altaffilmark{2}, Daniel J. Majaess\altaffilmark{1,3,4}\\ David J. Lane\altaffilmark{1,3}, Kathleen E. Moncrieff\altaffilmark{1}}
\email{turner@ap.smu.ca}

\altaffiltext{1}{Saint Mary's University, Halifax, Nova Scotia, Canada.}
\altaffiltext{2}{Astronomical Observatory, Odessa National University, Odessa, Ukraine.}
\altaffiltext{3}{The Abbey-Ridge Observatory, Stillwater Lake, Nova Scotia, Canada.}
\altaffiltext{4}{Visiting Astronomer, Dominion Astrophysical Observatory, Herzberg Institute of Astrophysics, National Research Council of Canada.}

\begin{abstract} 
New and existing photometry for the G0 Ia supergiant HD 18391 is analyzed in order to confirm the nature of the variablity previously detected in the star, which lies off the hot edge of the Cepheid instability strip. Small-amplitude variability at a level of $\Delta V = 0.016\pm0.002$ is indicated, with a period of $P=123^{\rm d}.04\pm0^{\rm d}.06$. A weaker second signal may be present at $P=177^{\rm d}.84\pm0^{\rm d}.18$ with $\Delta V = 0.007\pm0.002$, likely corresponding to fundamental mode pulsation if the primary signal represents overtone pulsation ($123.04/177.84 = 0.69$). The star, with a spectroscopic reddening of E$_{B-V} = 1.02\pm0.003$, is associated with heavily-reddened B-type stars in its immediate vicinity that appear to be outlying members of an anonymous young cluster centered $\sim10\arcmin$ to the west and $1661\pm73$ pc distant. The cluster has nuclear and coronal radii of $r_n=3.5\arcmin$ and $R_c=14\arcmin$, respectively, while the parameters for HD 18391 derived from membership in the cluster with its outlying B stars are consistent with those implied by its Cepheid-like pulsation, provided that it follows the semi-period-luminosity relation expected of such objects. Its inferred luminosity as a cluster member is $M_V=-7.76\pm0.10$, its age $(9\pm1) \times10^6$ years, and its evolutionary mass $\sim19 M_{\sun}$. HD 18391 is not a classical Cepheid, yet it follows the Cepheid period-luminosity relation closely, much like another Cepheid impostor, V810 Cen.
\end{abstract}

\keywords{techniques: photometric---stars: variables: Cepheids---open clusters and associations}

\section{Introduction}

The Cepheid instability strip is well recognized as the principal region of the H-R diagram occupied by luminous variable stars of spectral types F, G, and K. Less well known is the region blueward and at higher luminosity than the Cepheid instability strip occupied by a variety of stars, typically of luminosity classes Ia to Ib, exhibiting variability and microvariability at much smaller amplitudes \citep{mr72,ru78,rb82}. Such stars appear to obey relationships, as a function of spectral type or color, between their luminosities and semi-periods of variability \citep{bu78b,bual78} that, with the inclusion of amplitude as an additional parameter, are adequately explained by evolutionary models of massive stars that incorporate mass loss \citep{ma80}.

The G0 Ia supergiant HD 18391 \citep{nm52} has been used previously to characterize the cool end of the above relationships, tied to observations by \citet{ru78}. It is listed as a potential member of the double cluster h \& $\chi$ Persei \citep{sc84}, as well as Cas OB6 \citep{st72}. The star's broad spectral lines were remarked upon by \citet{re30} and \citet{gr70}, although that feature appears to be related to high luminosity in late-type stars \citep{im77}. Broad band {\it UBV} photoelectric photometry for the star was tabulated by \citet{ar66}, based upon 4 observations from Kitt Peak National Observatory. Additional {\it UBV} photometry is given by \citet{ha70}, but the evidence for variability lies in the observations of \citet{ru78}, in which the star was observed to increase gradually in brightness over an interval of 85 nights during the winter of 1972-73.

Unlike other stars defining the semi-period-luminosity-color relation for supergiants \citep{bu78b}, the semi-period of $\ge84$ days for HD 18391 was estimated from a single incomplete cycle, and is not well established. Our interest in HD 18391 arose from its spectroscopic parameters, which enabled its interstellar reddening to be estimated using a newly-developed technique \citep{ko07,ko08}. 

\begin{figure}[h]
\includegraphics[width=0.80\textwidth]{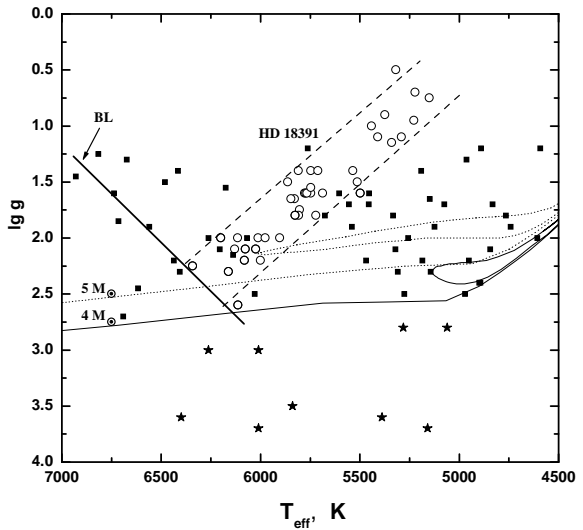}
\vspace{-0.8in}
\caption{\label{fig1} \small{The location of HD 18391 (numbered filled square) relative to other FGK giants and supergiants (other filled squares), known Cepheid variables (open circles), and the instability strip (dashed lines) in a HR diagram that plots surface gravity versus effective temperature. Evolutionary tracks are depicted for stars of 4 and 5 $M_{\sun}$, and ``BL'' denotes the edge of the blue loop evolutionary stage.}}
\end{figure}

The star's location in the spectroscopic H-R diagram (Fig.~\ref{fig1}) suggests a possible link to the Cepheid instability strip, which raises an intriguing question. Does the instability strip for Galactic Cepheids extend to pulsating GK supergiants like those in the Magellanic Clouds, where pulsation periods on the order of 100 days are not unusual? The present investigation was initiated in order to examine that question in greater detail.

\section{Observations and Air Mass Corrections}
The present study is based upon photometric observations of the brightness of HD 18391, as well as a detailed stellar atmosphere study of the star by \citet{ko07} and \citet{ko08} that generated the spectroscopic parameter $T_{\rm eff}=5775$ K using line ratio techniques, as well as $\log g =1.2$, metallicity [Fe/H] = 0.02, and microturbulent velocity 12.0 km s$^{-1}$. New CCD {\it V}-band differential photometry for HD 18391 was obtained on 101 nights between July 2007 and April 2009 using the 28-cm Schmidt-Cassegrain telescope of the semi-automated Abbey Ridge Observatory (ARO) located in Stillwater Lake, just outside of Halifax, Nova Scotia \citep[see][]{ma08}. All-sky CCD {\it BV} photometry for the same field was also obtained on the photometric nights of October 22, 23, and 26, 2007, using stars in the nearby open cluster NGC 225 as reference standards \citep{ho61}. Finally, we obtained CCD spectra in October 2008 for some luminous stars in the immediate vicinity of HD 18391, using the Cassegrain spectrograph on the Plaskett telescope of the Dominion Astrophysical Observatory, at a dispersion of 60 ${\rm \AA}$ mm$^{-1}$. 

\begin{deluxetable}{lllll}
\tabletypesize{\scriptsize}
\tablecaption{$BV$ magnitudes and colors for program objects} 
\label{tab1}
\tablewidth{0pt}
\tablehead{\colhead{Star} &\colhead{$V$} &\colhead{$B-V$}  &\colhead{Source}}
\startdata
HD 18391 &$6.937\pm0.004$ &$1.966\pm0.003$ &This paper \\
(G0 Ia) &$6.936\pm0.005$ &$1.982\pm0.015$ &Tycho$^1$ \\
&$6.890$ $\pm$ \ldots &$1.950$ $\pm$ \ldots & \citet{ar66} \\
&$6.890$ $\pm$ \ldots &$1.930$ $\pm$ \ldots & \citet{ha70} \\
\\
HD 18327 &$8.729\pm0.010$ &$0.222\pm0.013$ &This paper \\
(A0) &$8.732\pm0.011$ &$0.211\pm0.015$ &Tycho$^1$ \\
\\
BD+57$^{\circ}$668 &$9.797\pm0.001$ &$0.605\pm0.007$ &This paper  \\
(G0 V:) &$9.827\pm0.019$ &$0.676\pm0.031$ &Tycho$^1$ 
\enddata
\tablecomments{$^1$\citet{pe97}}
\end{deluxetable}

Differential photometry for HD 18391 was obtained relative to HD 18327 (A0), with BD+57$^{\circ}$668 (G0: V) serving as a check star. All-sky photometry for the three stars yielded the results summarized in Table \ref{tab1}, where comparison estimates for the magnitudes and colors are taken from mean Tycho {\it V}-band observations \citep{pe97}, \citet{ar66}, and \citet{ha70}. The mean Tycho {\it V} magnitude for HD 18391 was adjusted according to the precepts of \citet{be00}, but only half of the suggested correction to {\it B--V} color was used, otherwise the result would be significantly bluer than the values found by \citet{ar66}, \citet{ha70}, and this study. The mean Tycho values for HD 18327 and BD+57$^{\circ}$668, which are much bluer, are cited without adjustment.

Because the comparison stars are much bluer than HD 18391, it was necessary to adjust the differential photometry for air mass effects, atmospheric extinction being well known to diminish the brightness of blue stars by larger amounts than that of red stars as air mass increases. HD 18391 is a yellow star of very large reddening \citep{ko08}, so its brightness relative to bluer comparison stars would appear to increase artificially when viewed through large air masses if such corrections were not made. Given the low level of light variability in HD 18391, tests for such an effect are essential to confirm that any detected variability originates in the star rather than Earth's atmosphere.

\begin{figure}[ht]
\includegraphics[width=0.45\textwidth]{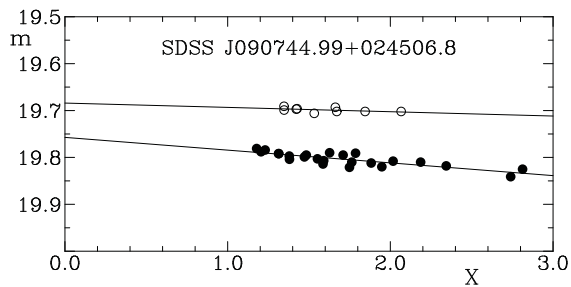}
\caption{\label{fig2} \small{The dependence of observations by \citet{fu06} for SDSS $J090744.99+024506.8$ on air mass X. Filled symbols represent {\it g}-band observations, open symbols {\it r}-band observations offset by $-0.1$. Straight lines are best fits to the observed trends, with residual scatter of $\pm0.008$ in {\it g} and $\pm0.005$ in {\it r}.}}
\end{figure}

Such an effect can be seen in differential photometry by \citet{fu06} from Mt. Hopkins for the high latitude hypervelocity blue star SDSS J090744.99+024506.8 (HVS) at $9^{\rm h}07^{\rm m}45^{\rm s}.0$, $+2^{\circ}45\arcmin07\arcsec$ (J2000). \citet{fu06} concluded that the star was a small amplitude variable because their differential CCD photometry relative to redder field stars in its vicinity indicated a brightness decrease on two nights of observation, coincident with an increase in the star's air mass. Temporal plots of the brightness of HVS on the two nights are replotted in Fig.~\ref{fig2} to display its changing brightness as a function of air mass X. For comparison stars with colors typical of G dwarfs relative to HVS, our experience with photometry from Kitt Peak suggests that the expected trends should have functional dependences of about $-0.03$ for {\it g}-band observations and $-0.01$ for {\it r}-band observations. The observed dependences are $-0.027$ and $-0.009$, respectively. The decreasing brightness of HVS as a function of increasing air mass is therefore exactly what is predicted for a blue star being compared with redder comparison stars, implying that its light variations are not intrinsic to the star.

The nonvariable status of HVS appears to be important to our understanding of the star, given that \citet{fu06} based their conclusion that it is a slowly-pulsating B-type main-sequence star (SPB) primarily on its purported variability. Since HVS is not variable, the original spectrum of the star obtained by \citet{br05} becomes a vital diagnostic tool. The spectrum displays broad hydrogen Balmer lines that appear to terminate near H13 or H14, which normally corresponds to a surface gravity of $\log g\simeq 4.5$ in B-type stars \citep[e.g.,][]{vi51}. That is typical of objects lying below the main sequence, such as the blue horizontal branch stars originally considered by \citet{br05}, and more recently by \citet{dv07}, as a potential origin for the object. Perhaps that possibly should be reconsidered, given that it is vital to any conclusions regarding the dynamical origin of hypervelocity stars?

\begin{figure}[ht]
\includegraphics[width=0.45\textwidth]{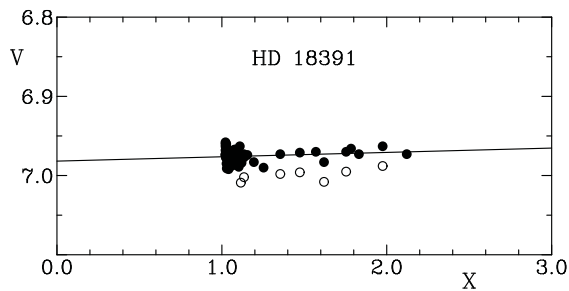}
\caption{\label{fig3} \small{The dependence of Geneva observations for HD 18391 on air mass X. Open symbols represent data from six nights of low quality, subsequently offset by $-0.025$ to match observations on other nights. The straight line is a best fit to the latter.}}
\end{figure}

The Geneva observations for HD 18391 \citet{ru78}, supplemented by a few additional unpublished observations by the group \citep{bu07}, were also checked for air mass effects, despite their origin from calibrated all-sky photometry. The site of the observations is not specified in the original references, but was assumed to be Gornergrat. The {\it V}-band observations are plotted in Fig.~\ref{fig3} as a function of inferred air mass, where it appears that a small zero-point offset affects observations on six of the nights, most of which were designated as being of low quality. An adjustment of $-0.025$ was adopted to bring them into agreement with the remaining observations, and the resulting data were adjusted for a small air mass term in the data (equivalent to an overly steep color dependence) and adjusted to a mean magnitude near $V=6.94$, to be consistent with the present data set.

\begin{figure}[h]
\includegraphics[width=0.45\textwidth]{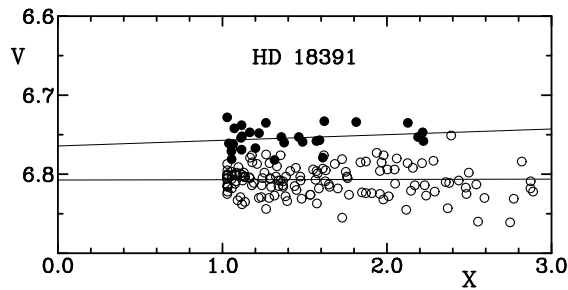}
\caption{\label{fig4} \small{The dependence of Abbey Ridge observations for HD 18391 on air mass X. Filled symbols represent observations from the first portion of the observing period, open symbols observations made after a change in CCD detectors. Straight lines are best fits to the observed trends.}}
\end{figure}

\begin{figure}[h]
\includegraphics[width=0.45\textwidth]{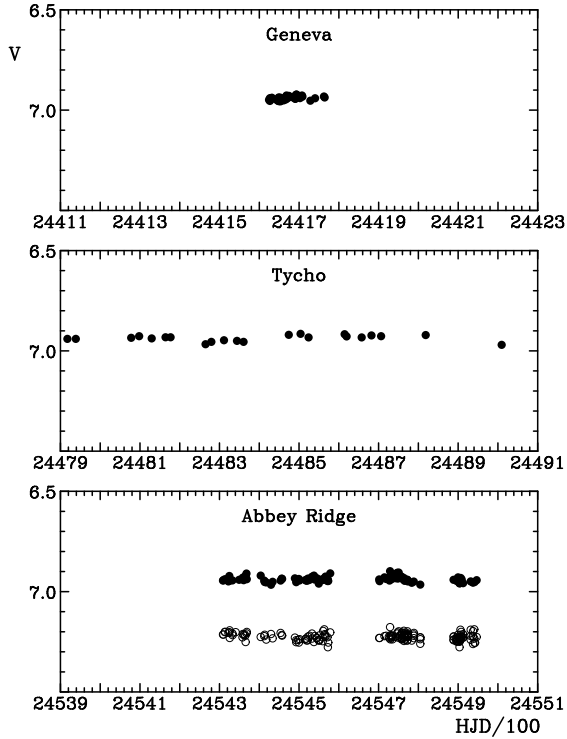}
\caption{\label{fig5} \small{Combined observations for HD 18391 from \citet{ru78} and \citet{bu07}, top, \citet{pe97}, middle, and this paper (bottom) plotted temporally. Open symbols in the bottom plot represent observations of the check star for the Abbey Ridge observations, offset by $-2.35$.}}
\end{figure}

A similar plot is presented in Fig.~\ref{fig4} for our raw differential photometry relative to HD 18327. A zero-point offset is present midway through our data set as a result of a change in CCD detectors, which produced a change in the color dependence for the observations. There is also a small difference in air mass dependence between observations in 2007 and 2008-09. Corrections for all such effects, an air mass dependence, zero-point offset, and differences relative to the results of our all-sky photometry, were made to all observations, with nightly means summarized in Table~\ref{tab2}, including a few combined observations from adjacent nights. Plots of the resulting time dependence for the Geneva, Tycho, and ARO observations are shown in Fig.~\ref{fig5}.

\begin{deluxetable}{cccccc}
\tabletypesize{\scriptsize}
\tablecaption{$V$-band observations of HD 18391 2007-2009} 
\label{tab2}
\tablewidth{0pt}
\tablehead{\colhead{HJD} &\colhead{$V$} &\colhead{HJD} &\colhead{$V$} &\colhead{HJD} &\colhead{$V$}}
\startdata
2454308.737 &6.945 &2454522.622 &6.935 &2454748.599 &6.935 \\
2454309.716 &6.943 &2454526.589 &6.941 &2454750.586 &6.905 \\
2454313.679 &6.938 &2454532.548 &6.926 &2454753.662 &6.924 \\
2454322.605 &6.949 &2454537.517 &6.920 &2454757.658 &6.925 \\
2454324.644 &6.922 &2454539.563 &6.940 &2454759.631 &6.936 \\
2454325.638 &6.945 &2454544.558 &6.933 &2454763.616 &6.932 \\
2454332.620 &6.944 &2454549.589 &6.960 &2454764.610 &6.943 \\
2454349.235 &6.941 &2454551.555 &6.940 &2454765.513 &6.946 \\
2454358.693 &6.930 &2454559.539 &6.939 &2454771.589 &6.941 \\
2454360.776 &6.943 &2454560.577 &6.942 &2454772.643 &6.941 \\
2454362.661 &6.932 &2454563.554 &6.934 &2454773.613 &6.949 \\
2454364.730 &6.925 &2454564.526 &6.935 &2454783.483 &6.956 \\
2454365.699 &6.921 &2454565.527 &6.935 &2454788.567 &6.950 \\
2454367.778 &6.910 &2454566.532 &6.925 &2454804.736 &6.965 \\
2454368.606 &6.938 &2454571.530 &6.947 &2454888.588 &6.941 \\
2454403.461 &6.920 &2454572.550 &6.940 &2454895.579 &6.939 \\
2454412.484 &6.945 &2454573.536 &6.947 &2454896.535 &6.947 \\
2454413.570 &6.952 &2454578.543 &6.909 &2454899.638 &6.929 \\
2454417.604 &6.954 &2454701.568 &6.939 &2454901.607 &6.941 \\
2454429.452 &6.966 &2454702.543 &6.948 &2454903.570 &6.959 \\
2454433.478 &6.951 &2454714.763 &6.929 &2454904.574 &6.958 \\
2454453.496 &6.945 &2454718.651 &6.937 &2454905.636 &6.940 \\
2454456.451 &6.935 &2454725.732 &6.932 &2454906.534 &6.932 \\
2454457.451 &6.937 &2454727.559 &6.945 &2454907.575 &6.944 \\
2454489.487 &6.934 &2454728.744 &6.898 &2454908.677 &6.939 \\
2454492.680 &6.953 &2454729.653 &6.929 &2454912.574 &6.958 \\
2454497.745 &6.940 &2454730.608 &6.932 &2454931.573 &6.949 \\
2454499.496 &6.937 &2454731.667 &6.915 &2454936.592 &6.953 \\
2454501.545 &6.944 &2454732.662 &6.938 &2454937.577 &6.955 \\
2454516.495 &6.942 &2454734.652 &6.928 &2454941.564 &6.952 \\
2454517.496 &6.944 &2454742.777 &6.933 &2454946.572 &6.943 \\
2454518.498 &6.945 &2454746.627 &6.906 & & \\
2454521.501 &6.945 &2454747.629 &6.909 & & 
\enddata
\end{deluxetable}

\section{Period Analysis}

It is not evident from the available observations that HD 18391 is a variable star. The standard deviations for the various data sets are $\pm0.009$ for Geneva observations, $\pm0.015$ for Tycho observations, and $\pm0.012$ for ARO data in Table \ref{tab2}, $\pm0.014$ for the individual observations. The dispersion in the ARO check star observations is $\pm0.016$, which is larger than the scatter observed for HD 18391 only because the latter is 3 magnitudes brighter than the check star. Its signal therefore had a higher signal-to-noise ratio (SNR) than those for the comparison stars, and was affected less by photon statistics. Small SNRs for the comparison stars were inevitable, given that exposure times were kept short to avoid saturation of the target. The scatter for the check star also appears to be random, whereas the observations for HD 18391 display slight temporal trends. Similar observations for other stars observed at the ARO of comparable brightness to HD 18391 exhibit smaller dispersion, typically about $\pm0.008$, but the main conclusion is that any variability in HD 18391 is at best marginally detectable in existing observations. Previous conclusions by \citet{ru78} regarding the variability of HD 18391 may therefore have been affected by the small air mass effect noted earlier.

Despite uncertainties about the quality of the different data sets, the observations were analyzed using various routines available in the {\sc Peranso} software package of \citet{va06}, which includes the CLEANest routine \citep{fo95}. Fifteen different period finding routines were used with the combined observations, with the average value for the most significant period from twelve of them being $P=123^{\rm d}.04\pm0^{\rm d}.06$ s.d. A similar analysis of observations for the ARO check star, BD+57$^{\circ}$668, was made as a consistency check, yielding a dominant period at $P\simeq102$ days but without producing a realistic light curve.

\begin{figure}[h]
\includegraphics[width=0.40\textwidth]{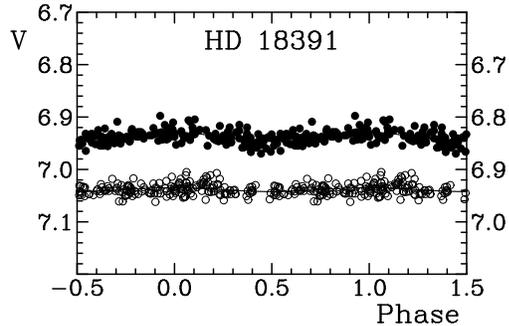}
\caption{\label{fig6} \small{The resulting light curve for HD 18391 (filled circles) from the adopted ephemeris for $P=123^{\rm d}.04$ (upper), along with the best-fitting sine wave. Open circles represent the prewhitened observations phased to $P=177^{\rm d}.84$ relative to the right-hand scale.}}
\end{figure}

The resulting light curve for HD 18391 fits an ephemeris given by:
\begin{equation}
\label{eqn1}
{\rm HJD}_{\rm max}=2441695.497+123.04\;E \;, 
\end{equation}
where {\it E} is the number of elapsed cycles. It is displayed in Fig.~\ref{fig6}. The inferred light amplitude is small, $\Delta V=0.016\pm0.002$, comparable to the scatter in the observations and marginally significant, and perhaps unusual given that pulsation amplitude typically increases with rate of mass loss and luminosity in variable supergiants \citep{ma80}. The lack of significant light variability for HD 18391 presumably implies a small rate of mass loss or strong pulsational damping by convection.

A second signal appears in the data at $P=177^{\rm d}.84\pm0^{\rm d}.18$ s.d. after removal of the primary signal. The signal is weak, but may be real, given that it corresponds with expectations for fundamental mode pulsation if the $123^{\rm d}$ signal originates from overtone pulsation ($123.04/177.84 = 0.69$). The ephemeris for the second signal is:
\begin{equation}
{\rm HJD}_{\rm max}=2441735.442+177.84\;E \;, 
\end{equation}
with a light amplitude of $\Delta V=0.007\pm0.002$. The light curve for the prewhitened observations with that period is shown in the lower portion of Fig.~\ref{fig6}.

Conceivably some of the residual scatter in the light curves of Fig.~\ref{fig6} results from additional periodicities in the observations, temporal changes in light amplitude, or to rapid period changes in the star's pulsation. HD 18391 displays abundance patterns typical of stars evolving towards a first crossing of the Cepheid instability strip, including a strong lithium feature \citep{ko05}. As such, radial pulsation in the star should undergo period increases of several hours per year \citep[e.g.,][]{tu06a}, complicating the analysis significantly. The presence of a second period is remarkably similar to what is observed in V810 Cen, a F8 Ia supergiant similar in nature to HD 18391 pulsating with periods of 153 and 104 days and displaying temporal changes in light amplitude \citep{ke98,bu06}. The light amplitude of HD 18391 is an order of magnitude smaller than that of V810 Cen, so additional observations and long-term monitoring are crucial to search for similar effects, as well as to confirm the small-amplitude variability implied by observations to date.

\section{Further Consistency Checks}

\begin{figure}[h]
\includegraphics[width=0.40\textwidth]{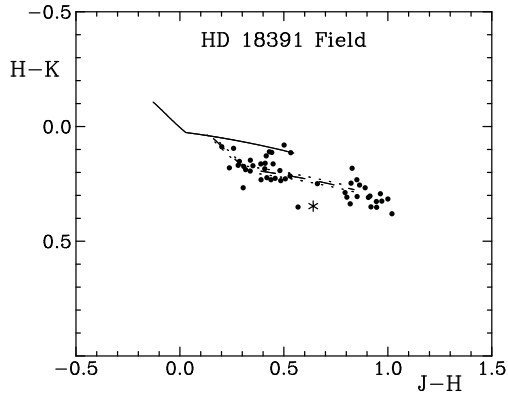}
\caption{\label{fig7} \small{A 2MASS color-color diagram, {\it H--K} versus {\it J--H}, for stars within $5\arcmin$ of HD 18391 having magnitude uncertainties less than $\pm0.03$. The intrinsic relation for main-sequence stars is plotted as a solid line, and the same relation reddened by {\it E(J--H)} $=0.29\pm0.03$ as a dashed line surrounded by dotted lines. The kink corresponds to A0 stars. HD 18391 is depicted by an asterisk.}}
\end{figure}

Independent checks can be made into the nature of the star. For example, \citet{ko08} derive a spectroscopic reddening for HD 18391 of {\it E(B--V)} $=0.991$, but that is based upon a different adopted {\it B--V} color. With the results of Table~\ref{tab1}, the reddening becomes {\it E(B--V)} $=1.020\pm0.003$. A field reddening for HD 18391 can be found using stars in its immediate vicinity. The observed colors of stars in the 2MASS catalogue \citep{cu03} lying within $5\arcmin$ of the star are shown in Fig.~\ref{fig7}. When analyzed using the method illustrated by \citet{tu08}, the data imply a field reddening of {\it E(J--H)} $=0.29\pm0.03$, which corresponds to {\it E(B--V)} $=1.04\pm0.09$, consistent with the spectroscopic result. The obvious A-type main-sequence stars in the sample have inferred distances of $\sim650$ pc, implying that the reddening in the field originates relatively nearby, certainly foreground to the G0 Ia supergiant.

\begin{deluxetable}{llrrccc}
\tabletypesize{\scriptsize}
\tablecaption{Luminous stars within $\sim15\arcmin$ of HD 18391} 
\label{tab3}
\tablewidth{0pt}
\tablehead{\colhead{Star} &\colhead{Sp.T.} &\colhead{$V$} &\colhead{$B-V$} &\colhead{Source} &\colhead{E(B-V)} &\colhead{$V_0$-$M_V$} }
\startdata
HD 18391 &G0 Ia &6.94 &1.97 &1 &1.02 &\ldots \\
LS+57$^\circ$111 &OB+ce,h &11.42 &0.85 &2 &\ldots &\ldots \\
LS+57$^\circ$113 &B0.5 V &11.03 &0.81 &1,2 &1.09 &11.13 \\
LS+57$^\circ$114 &A2 Ib &11.31 &1.00 &2 &0.95 &13.18 \\
LS+57$^\circ$115 &B1.5 Vn &12.46 &0.83 &1,2 &1.08 &11.48 \\
3709 490 1 &B2 Vnn &11.34 &0.60 &1,4 &0.84 &10.83 \\
LS+57$^\circ$116 &b1 v &11.94 &0.89 &2 &1.16 &11.21 \\
LS+57$^\circ$117 &B0 V &10.82 &0.87 &1,2 &1.17 &10.77 \\
V637 Cas &M2 Iab-Ib &9.88 &2.70 &3 &0.99 &11.31 \\
Mean & & & & & &11.12 \\
s.e. & & & & & &$\pm0.12$
\enddata
\tablecomments{Sources: 1. This paper, 2. \citet{ha70}, 3. \citet{le70}, 4. \citet{pe97}}
\end{deluxetable}

\begin{figure}[h]
\includegraphics[width=0.40\textwidth]{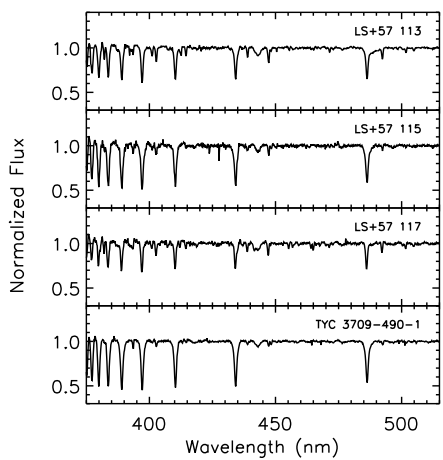}
\caption{\label{fig8} \small{Low-dispersion spectra for stars near HD 18391.}}
\end{figure}

The distance to HD 18391 can be estimated using luminous stars located in close proximity. Table~\ref{tab3} lists early-type and luminous stars located within $\sim15\arcmin$ of the G0 Ia supergiant from Volume I of {\it Luminous Stars in the Northern Milky Way} \citep{ha59}, for which photometric observations are available from \citet{ha70}, \citet{le70}, and \citet{pe97}. Included is the LC class variable M supergiant V637 Cas. Spectral types are available for most stars from the ADS database, including results from \citet{ww78} for V637 Cas, but our spectroscopic observations of the B-type stars, presented in Fig.~\ref{fig8}, were used to test and augment them. The resulting spectral classifications produce the spectroscopic distance moduli given in the last column of Table~\ref{tab3}, where a value of $R=A_V/E(B-V) = 3.2$ has been adopted for the ratio of total-to-selective absorption in this region of the Galaxy \citep{tu76}.

\begin{figure}[ht]
\begin{center}
\includegraphics[width=0.40\textwidth]{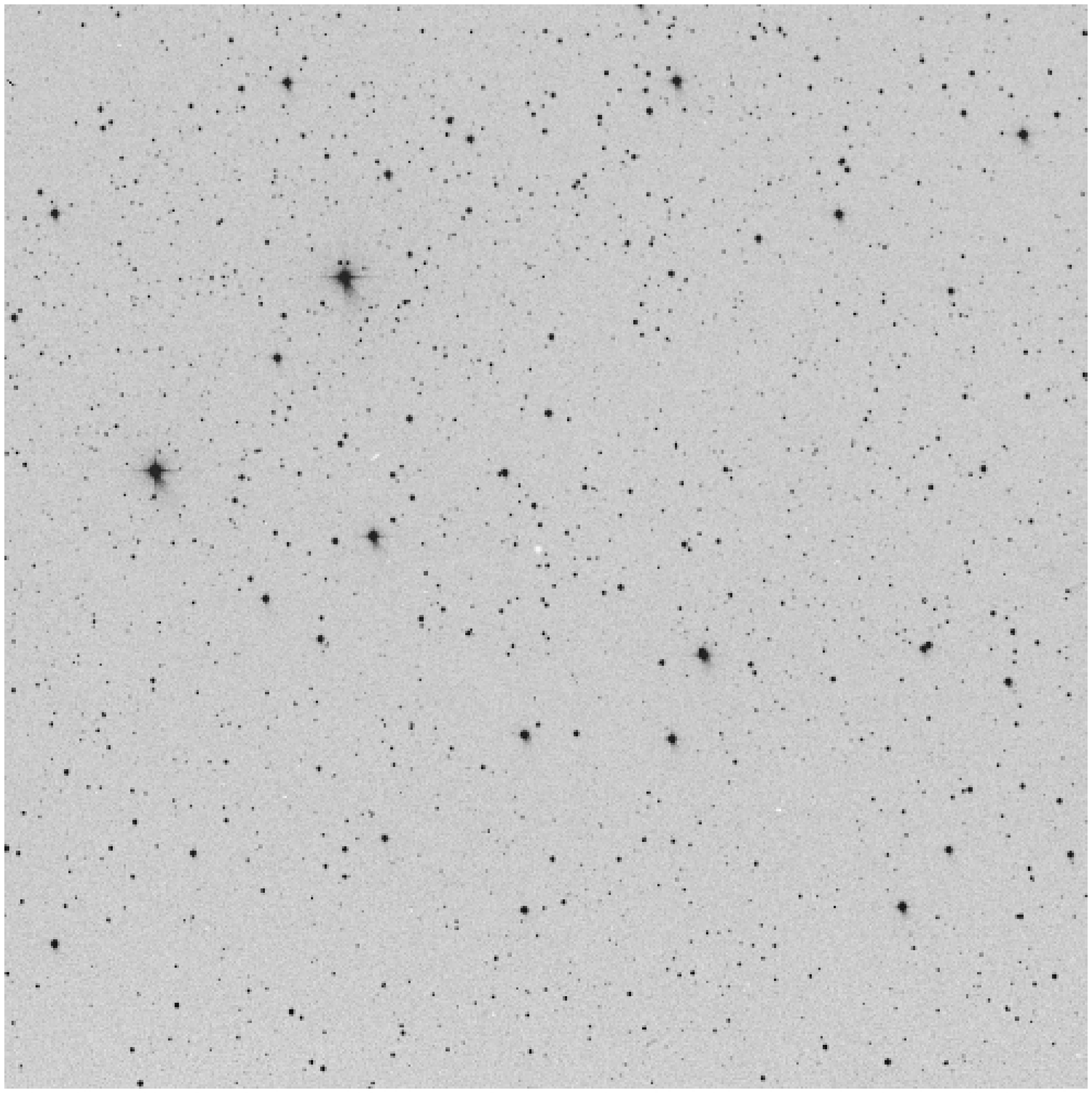}
\caption{\label{fig9} \small{The view of the anonymous cluster located at 2:58:28, +57:37:30 (J2000) on the Palomar Observatory Sky Survey blue plate. The field measures $30\arcmin$ by $30\arcmin$ with north up, and HD 18391 is the bright star on the left edge of the field. [The National Geographic Society-Palomar Observatory Sky Atlas (POSS-1) was made by the California Institute of Technology with grants from the National Geographic Society.]}}
\end{center}
\end{figure}

\begin{figure}[ht]
\includegraphics[width=0.40\textwidth]{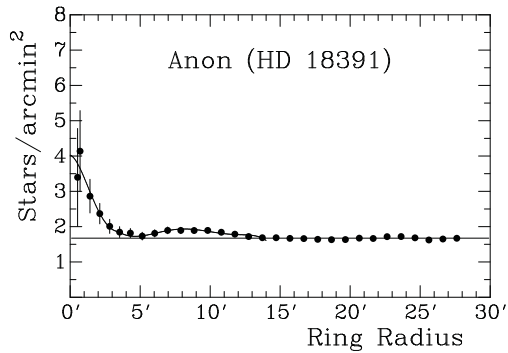}
\caption{\label{fig10} \small{Ring counts for $J\le15$, with their Poisson uncertainties, for the anonymous cluster centered at 2:58:28, +57:37:30 (J2000). The field star level based upon rings with $r\ge13\arcmin$ is shown, as is a possible trend for cluster stars. The dimensional parameters are $r_n=3.5\arcmin$ and $R_c=14\arcmin$.}}
\end{figure}

All of the luminous stars near HD 18391 are reddened by $E(B-V)\simeq 1.0$, with distance moduli of $\sim11$. The exception is the A2 Ib supergiant LS+57$^\circ$114, which does not fit coevally with the remainder of the group. The mean unreddened distance modulus for the group is $V_0-M_V=11.12\pm0.12$, corresponding to a distance of $1678\pm96$ pc. Interestingly enough, there is a sparse group of early-type stars located $\sim10\arcmin$ west of HD 18391 at J2000 co-ordinates 2:58:28, +57:37:30 (Fig.~\ref{fig9}). The cluster is only marginally detectable by eye, although similarly sparse groups have been detected previously in such fashion by the lead author \citep[e.g.,][]{tu85,tu93}. Visually, the cluster displays the characteristics of a loose group in the terminal stages of dissolution into the field, as confirmed by star counts. Ring counts about the adopted cluster center for $J\le15$ are shown in Fig.~\ref{fig10}, from which one derives nuclear and coronal radii of $r_n=3.5\arcmin$ (1.7 pc) and $R_c= 14\arcmin$ (6.8 pc), respectively, in the nomenclature of \citet{kh69}. An unusual feature is the increase in star densities in the cluster corona, indicative of a ringlike feature. A similar characteristic is shared by Collinder 70, better known as the Orion Ring, a young cluster of stars surrounding $\epsilon$ Ori in Orion's Belt that is in the advanced stages of dissolution into the field.

\begin{figure}[ht]
\includegraphics[width=0.35\textwidth]{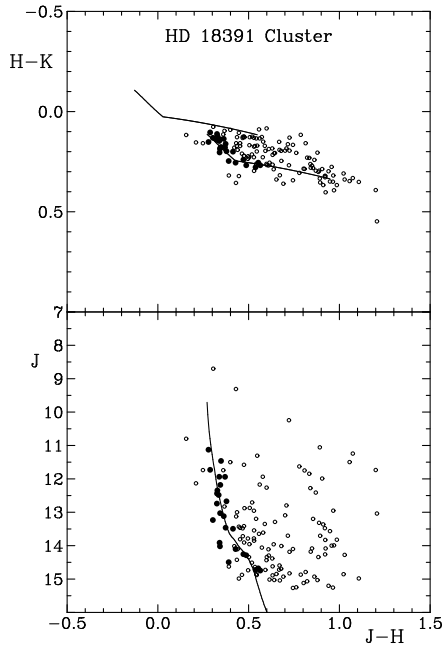}
\caption{\label{fig11} \small{A 2MASS color-color diagram, {\it H--K} versus {\it J--H} (upper section), for stars within $5\arcmin$ of 2:58:28, +57:37:30 (J2000) having magnitude uncertainties less than $\pm0.05$. The intrinsic relation for main-sequence stars is plotted as a solid line, as well as for {\it E(J--H)} $=0.40$. Filled circles represent likely cluster members, open circles probable field stars. Note the large number of reddened stars blueward of the A0 star kink. The lower section is a color-magnitude diagram for the same stars, indicating the best zero-age main-sequence fit for {\it J--M}$_J=11.07\pm0.15$.}}
\end{figure}

The parameters for this group of stars, most of which can be identified as early B-type dwarfs from their 2MASS colors (Fig. \ref{fig11}), are similar to those for stars lying closer to HD 18391. The implied reddening is {\it E(J--H)} $=0.40\pm0.02$, corresponding to {\it E(B--V)} $=1.43\pm0.07$, and the distance modulus is $J_0-M_J=11.07\pm0.15$, corresponding to a distance of $1638\pm113$ pc. The adopted mean of the two estimates is $d=1661\pm73$ pc ($11.10\pm0.10$ in intrinsic distance modulus). HD 18391 and most of the luminous stars of Table~\ref{tab3} lie in the cluster corona, a by-now common characteristic of many young star clusters \citep[e.g.,][]{bu78a}. The predicted number of cluster members for $J\le15$ lying within $3.5\arcmin$ of the cluster center is $16\pm8$ according to the star counts, compared with exactly 16 identified in this region from 2MASS photometry. The total number of cluster members with $J\le15$ according to the star counts is $102\pm35$, typical of a poorly-populated cluster.

\begin{deluxetable}{lcc}
\tabletypesize{\scriptsize}
\tablecaption{Radial Velocity Data.} 
\label{tab4}
\tablewidth{0pt}
\tablehead{\colhead{Star} &\colhead{JD} &\colhead{V$_{\rm R}$(km s$^{-1}$)}}
\startdata
HD 18391 &2449739.3755 &$-41.3$ \\
HD 18391 &2452711.3171 &$-44.4$ \\
HD 18391 &2453785.2410 &$-43.5$ 
\enddata
\end{deluxetable}

\begin{figure}[ht]
\includegraphics[width=0.40\textwidth]{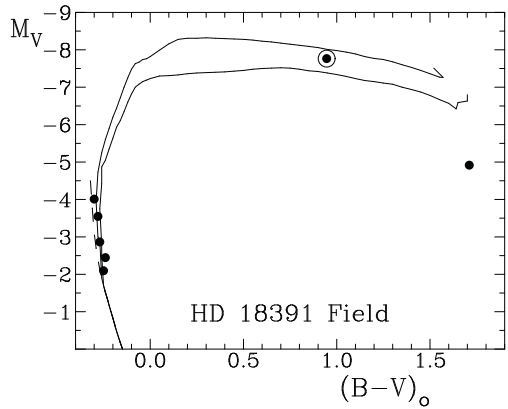}
\caption{\label{fig12} \small{An extinction-free H-R diagram for neighboring stars to HD 18391, the circled point. Isochrones are shown for $\log t = 6.9$ (upper) and $\log t = 7.0$ (lower).}}
\end{figure}

If HD 18391 is a member of the cluster and B-star group, its implied luminosity is $M_V=-7.76\pm0.10$. It appears to fit in an evolutionary sense as well, according to the unreddened color-magnitude diagram of Fig.~\ref{fig12} for stars in the B-star group. The stars appear to have an age of $\log t = 6.95\pm0.05$, or $(9\pm1) \times 10^6$ years, according to an isochrone fit to models by \citet{me93}. The irregular M supergiant variable V637 Cas does not match the same isochrone very well, almost certainly because of heavy mass loss during its previous evolution \citep[e.g.,][]{jk90}. A similar situation applies to the M supergiant BC Cyg as a member of Berkeley 87 \citep{tf82,tu06b,ma08}, and the M3 Iab-Ib member of NGC 2439 \citep{tu77}. The mass of stars at the red turnoff (RTO) for a $9 \times 10^6$ year-old cluster is $18\frac{1}{2} M_{\sun}$ according to \citet{me93}, implying that the mass of HD 18391 is $\sim19 M_{\sun}$. Possible membership of HD 18391 in Cas OB6 can be ruled out. Cas OB6 members include evolved O4 and O5 stars \citep{hi06}, implying a much younger age, and are more distant: 2.01 kpc according to \citet{me95}, $\sim1.9$ kpc as adopted by \citet{hi06}. They also have radial velocities of $\sim-50$ km s$^{-1}$, marginally more negative than the measured velocities for HD 18391 summarized in Table~\ref{tab4}. Attempts were made to measure radial velocities for the associated B stars observed spectroscopically in Fig.~\ref{fig8}, but with negative results because of zero-point problems. The three coolest stars in the sample were found to have nearly identical velocities, however.  

\begin{figure}[ht]
\includegraphics[width=0.45\textwidth]{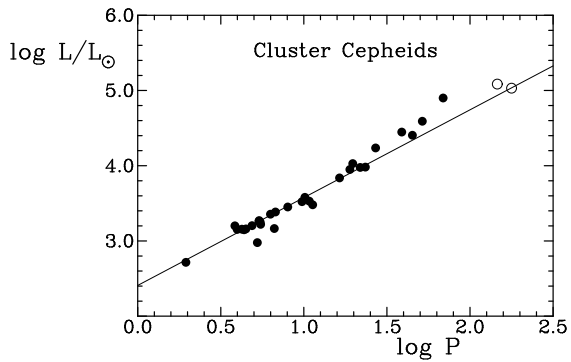}
\caption{\label{fig13} \small{The period-luminosity relation for Galactic Cepheids associated with open clusters and associations. Classical Cepheids are denoted by filled circles, and the Cepheid-like supergiants V810 Cen and HD 18391 by open circles.}}
\end{figure}

The implied luminosity of HD 18391, $M_V=-7.76$, or $M_{\rm bol}=-7.79$, agrees closely with expectations from published semi-period-luminosity relations for Cepheid-like supergiants \citep{bu78b}, provided that the longer period ($P=178^{\rm d}$) applies. Interestingly enough, both HD 18391 and V810 Cen also fit predictions from the Cepheid period-luminosity relation, as shown in Fig.~\ref{fig13}. The results for the two FG class Ia supergiants are based upon their assumed membership in Anon (HD 18391) and Stock 14 \citep[see][]{tu82}, although the case for V810 Cen as a cluster member has been questioned on the basis of a conflict with the radial velocity of Stock 14 stars \citep{ke98}. The case for HD 18391 also requires radial velocity confirmation.

The radius of HD 18391 implied by its luminosity and effective temperature is $329 R_{\sun}$, whereas a value of $583 R_{\sun}$ would characterize a Cepheid of the same pulsation period \citep{tb02}. The implied radius also produces a value of $\log g=0.69$ with the effective temperature inferred from the model atmosphere analysis \citep{ko08}, consistent with the cited value of $\log g= 1.2$ when non-LTE and possible mass-loss effects are taken into account. But HD 18391 is clearly not a Cepheid. Its variability presumably originates from pulsation, radial pulsation if the implied periods are correct, highly damped by convection. Variability possibly linked to the rotation of HD 18391 would imply a rotation rate of $\sim90$ km s$^{-1}$, much too large for a star that has expanded by two orders of magnitude since the main sequence stage. The color variations of the star are also in antiphase to the brightness variations \citep{bu78b}, further constraining the nature of pulsation in the star \citep{ma80}.

\section{Discussion}
The suspected, long-term, low-amplitude, light variability of HD 18391 is confirmed by recent observations of the star, although the light curve derived from recent CCD observations contains considerable scatter, a problem that may be associated with the difficulty of making accurate extinction corrections for observations from a site located near sea level. The dominant period of variability of $P=123^{\rm d}.04$ is reasonably solid, but implies a semi-period of $61^{\rm d}.52$, smaller than the value of $\ge84^{\rm d}$ originally estimated by \citet{ru78}. The latter value lies closer to $P=177^{\rm d}.84$ (semi-period of $88^{\rm d}.92$) found for the weaker second signal. A luminosity of $M_V=-7.76$ is implied for HD 18391 from likely membership in a surrounding group of luminous stars that, with the G0 Ia supergiant, are outliers to an anonymous, poorly-populated open cluster $\sim10\arcmin$ to the west that appears to be dissolving into the field. The likely age of the B-star subgroup surrounding the G0 Ia supergiant is $\log t = 6.95\pm0.05$, or $(9\pm1) \times 10^6$ years. The luminosity estimate is consistent with the existing semi-period-luminosity relation of \citet{bu78b} for such objects.

It is interesting to note that both V810 Cen and HD 18391 fit the period-luminosity relation for cluster and association Cepheids extremely well, in fact unerringly in the case of HD 18391, despite the fact that their implied effective temperatures and luminosities place them well to the hot side of the classical Cepheid instability strip. The mechanism responsible for the variability of such Cepheid impostors is still poorly constrained, although non-radial pulsation and effects tied to mass loss have been considered previously \citep{ma80}. The present results for HD 18391 suggest the possibility of radial pulsation. It is curious how well Cepheid-like supergiants obey the classical period-luminosity relation for Cepheids. It is not clear why that should be, although their presence among the luminous population of galaxies may be important to the identification and use of Cepheids as distance indicators, particularly for low-amplitude variables.

\section*{Acknowledgements}
The authors are grateful to Gilbert Burki for kindly providing unpublished observations of HD 18391 that helped to extend the time baseline of observations for the star.

\end{document}